\documentclass[final, 5p, times,  twocolumn]{elsarticle}

\usepackage{float}     

\usepackage{listings}
\usepackage{xcolor} 
\usepackage{pgfplots}
\usepackage[most]{tcolorbox}
\usepackage{lipsum} \usepackage[most]{tcolorbox}

\definecolor{codegreen}{rgb}{0,0.6,0}
\definecolor{codegray}{rgb}{0.5,0.5,0.5}
\definecolor{codepurple}{rgb}{0.58,0,0.82}
\definecolor{backcolour}{rgb}{0.95,0.95,0.92}

\lstdefinestyle{mystyle}{
 backgroundcolor=\color{backcolour},  commentstyle=\color{codegreen},
 keywordstyle=\color{magenta},
 numberstyle=\tiny\color{codegray},
 stringstyle=\color{codepurple},
 basicstyle=\ttfamily\footnotesize,
 breakatwhitespace=false,     
 breaklines=true,         
 captionpos=b,          
 keepspaces=true,         
 numbers=left,          
 numbersep=5pt,         
 showspaces=false,        
 showstringspaces=false,
 showtabs=false,         
 tabsize=2
}

\usepackage{graphicx}

\usepackage{algorithm2e}
\usepackage{multirow}
\usepackage{booktabs}
\usepackage{amsmath,amssymb,amsfonts}
\usepackage{algorithmic}
\usepackage{graphicx}
\usepackage{textcomp}
\usepackage{booktabs}
\usepackage{rotating}
\usepackage{caption}
\usepackage{multirow}
\usepackage[most]{tcolorbox}
\usepackage{url}

\usepackage{amsmath}

\usepackage{comment}
\usepackage{float}
\usepackage{xcolor}
\usepackage{diagbox}
\usepackage{epstopdf}
\usepackage{caption}
\usepackage{subfig}

\usepackage{hyperref}
\usepackage{blindtext}
\usepackage{pifont}\usepackage{todonotes}
\newcommand{\gl}[1]{\todo[inline]{GL: #1}}

\usepackage{xcolor}
\def\BibTeX{{\rm B\kern-.05em{\sc i\kern-.025em b}\kern-.08em
    T\kern-.1667em\lower.7ex\hbox{E}\kern-.125emX}}
    \pgfplotsset{compat=1.18} 

\lstdefinestyle{mystyle}{
 backgroundcolor=\color{backcolour},  commentstyle=\color{codegreen},
 keywordstyle=\color{magenta},
 numberstyle=\tiny\color{codegray},
 stringstyle=\color{codepurple},
 basicstyle=\ttfamily\footnotesize,
 breakatwhitespace=false,     
 breaklines=true,         
 captionpos=b,          
 keepspaces=true,         
 numbers=left,          
 numbersep=5pt,         
 showspaces=false,        
 showstringspaces=false,
 showtabs=false,         
 tabsize=2
}

\begin{document}

\begin{frontmatter}

\title{Sóley: Identification and Automated Detection of Logic Vulnerabilities in Ethereum Smart Contracts Using Large Language Models}

\author{Majd Soud,
Waltteri Nuutinen,
and Grischa Liebel
}\ead{majd18@ru.is, grischal@ru.is}

\address{Department of Computer Science, Reykjavik University, Reykjavik, Iceland}

\begin{abstract}
\textbf{Context:} Modern blockchain, such as Ethereum, supports the deployment and execution of so-called smart contracts, autonomous digital programs with significant value of cryptocurrency. Executing smart contracts requires gas costs paid by users, which define the limits of the contract's execution. 
Logic vulnerabilities in smart contracts can lead to excessive gas consumption, financial losses, and are often the root cause of high-impact cyberattacks.\\
\textbf{Objective:} Our objective is threefold: (i) empirically investigate logic vulnerabilities in real-world smart contracts extracted from code changes on GitHub, (ii) introduce Sóley, an automated method for detecting logic vulnerabilities in smart contracts, leveraging Large Language Models (LLMs), and (iii) examine mitigation strategies employed by smart contract developers to address these vulnerabilities in real-world scenarios.\\  
\textbf{Method:} We obtained smart contracts and related code changes from GitHub. To address the first and third objectives, we qualitatively investigated available logic vulnerabilities using an open coding method. We identified these vulnerabilities and their mitigation strategies. For the second objective, we extracted various logic vulnerabilities, applied preprocessing techniques, and implemented and trained the proposed Sóley model. We evaluated Sóley along with the performance of various LLMs and compared the results with the state-of-the-art baseline on the task of logic vulnerability detection.\\
\textbf{Results:} From our analysis of code changes on GitHub, we identified nine novel logic vulnerabilities, extending existing taxonomies with these vulnerabilities. Furthermore, we introduced several mitigation strategies extracted from observed developer modifications in real-world scenarios. Our Sóley method outperforms existing methods in automatically identifying logic vulnerabilities. Interestingly, the efficacy of LLMs in this task was evident without requiring extensive feature engineering.\\
\textbf{Conclusion:} Early identification of logic vulnerabilities from code changes can provide valuable insights into their detection and mitigation. Recent advancements, such as LLMs, show promise in detecting logic vulnerabilities and contributing to smart contract security and sustainability.

\end{abstract}


\end{frontmatter}

 \section{Introduction}
Smart contracts are self-executing programs and key components of modern blockchain such as Ethereum~\cite{buterin2014next}. Smart Contracts have gained widespread attention for their ability to enforce agreements without intermediaries, offering a decentralized and trustworthy approach~\cite{nakamoto2008bitcoin}.
The characteristics of smart contracts, including high financial value, immutability, and decentralized nature, have facilitated the growth of decentralized applications (DApps) that offer various functionalities in fields such as gaming, e-commerce, and more~\cite{wood2014ethereum}.
Smart contracts are typically implemented using high-level languages such as Solidity~\footnote{https://docs.soliditylang.org/en/v0.8.23/}. Their implementation consists of functions that define a sequence of instructions and a set of state variables~\cite{antonopoulos2018mastering}. 

Ethereum, as the first blockchain supporting smart contracts~\cite{buterin2014next}, has garnered significant developer interest, leading to the deployment of over 64 million smart contracts~\footnote{https://etherscan.io/contractsVerified} and a market capitalization exceeding \$300 billion~\footnote{https://etherscan.io/}.
On Ethereum, executing a smart contract requires fees known as \emph{``gas"}, a unit that measures the consumption of storage and computing resources~\cite{wood2014ethereum}. These fees, paid in Ether~\footnote{Ethereum cryptocurrency}, compensate for computing resources used during contract execution~\footnote{https://ethereum.org/developers/docs/gas}. Smart contracts with logic vulnerabilities are susceptible to severe consequences, including high-impact cyberattacks, execution failures due to excessive gas usage, wasted gas, and significant financial losses~\cite{chaliasos2024smart}. Identifying logic vulnerabilities is crucial for maintaining Solidity smart contracts as emphasized by Chaliasos et al.~\cite{chaliasos2024smart}. In their study, they highlight the inefficiency of existing smart contract security tools in detecting logic-related vulnerabilities, sanity checks, and logic errors. Logic vulnerabilities become even more critical when they exist in smart contract functions responsible for transferring Ether out of the contract, as this could potentially prevent the contract owners or users from accessing their money (i.e, Ether). A notable example is the Governmental contract~\cite{GovernMental}, where \$2.5 million worth of Ether got locked out. Therefore, ensuring a sustainable and secure development of smart contracts is imperative, and addressing logic vulnerabilities plays a crucial role in achieving this goal. Smart contracts, with their sequential code structure and the interrelation of statements, share similarities with natural language text. Moreover, they often conform to predetermined templates and standards, leading to recognizable patterns and recurring code structures~\cite{clack2016smart}. Consequently, Large Language Models (LLMs) and other Natural Language Processing (NLP) methods offer the potential for automating the detection of vulnerabilities and generating comments within smart contracts~\cite{hu2023large, zhao2024automatic}.
 
Motivated by the advancements achieved by LLMs in conventional software and the inherent characteristics of smart contract code, 
this paper aims to answer the following research questions. 
\begin{itemize}
    
\item  \textbf{RQ1 - (Identification)}: To what extent do historical code changes reveal logic vulnerabilities in smart contracts?
  
\item  \textbf{RQ2 - (Detection):}  How can we automatically detect logic vulnerabilities in smart contracts via LLMs?

\item  \textbf{RQ3 - (Mitigation):} What specific strategies do developers employ in their code changes to mitigate potential logic vulnerabilities in smart contracts?
\end{itemize}

With these RQs in mind, our methodology consists of several key steps. First, we gathered instances of smart contract vulnerabilities and their corresponding fixes from real-world Solidity code changes on GitHub. We then investigated the identified logic vulnerabilities using open coding~\cite{service2009book}, defining the available vulnerabilities and corresponding mitigation strategies. Subsequently, we developed Sóley to effectively detect logic vulnerabilities using LLMs without extensive feature engineering. Finally, we investigate the performance of various LLMs on the logic vulnerability  detection task.

Our work makes the following contributions:
\begin{itemize}
    \item \textbf{Sóley Development and Experiment.} We present Sóley, based on CodeBERTa, to detect logic-related vulnerabilities in smart contracts. We also explore the performance of well-known LLMs on the same task for various code vulnerabilities in smart contracts.

   \item  \textbf{Dataset.} We curated a large dataset of Solidity smart contracts labeled with vulnerability types, locations in the contracts, and other metadata. This dataset contains 50k contracts, 428,569 instances of code vulnerabilities, and 171,180 logic related vulnerabilities.

\item \textbf{Analysis of Logic code vulnerabilities and mitigation.} We conduct an in-depth analysis of logic vulnerabilities found in code changes within smart contracts and in literature, identifying nine novel vulnerabilities. Moreover, we introduce 15 mitigation strategies used by developers to address logic vulnerabilities.
\end{itemize}

We have made the models, scripts, and data utilized in this study openly available
to promote research replication and facilitate additional investigations by other researchers in the field.

The rest of the paper is organized as follows. In Section~\ref{sec:preliminaries}, we provide background information about smart contracts and the Ethereum blockchain. Our research methodology is outlined in Section~\ref{sec:method}. Section~\ref{sec:setup} presents the experiment evaluation and setup, while Section~\ref{sec:results} details our findings. Section~\ref{sec:empirical} provides discussions and implications. Section~\ref{sec:related_work} presents the related work. Additionally, we address threats to the validity of our work in Section~\ref{sec:threats} before concluding in Section~\ref{sec:conclusion}. \section{Background}
\label{sec:preliminaries}

\subsection{Blockchain and Ethereum}
Blockchains are transparent, decentralized ledgers of transactions~\cite{zheng2018blockchain}. A blockchain consists of a set of blocks, each containing a sequence of transactions that cannot be altered without affecting all subsequent blocks. Ethereum is a blockchain technology featuring a single virtual machine, the Ethereum Virtual Machine (EVM)~\cite{wood2014ethereum}. Ethereum provides users with two types of accounts: a smart contract-controlled account or an externally owned account (EOA)~\cite{ethereum2023accounts}. The latter allows users to initiate a transaction on the EVM to change its states or transfer cryptocurrency (i.e., Ether). Each transaction on the EVM consists of several attributes, such as a transaction hash, sender address, and receiver address.  All initiated transactions are added to the blockchain after verification based on cryptographic mechanisms to the blockchain.
\subsection{Smart contract and Solidity}
Smart contracts are self-executing digital agreements, programmed with high-level languages, that require deployment on the blockchain to execute~\cite{szabo1996smart, wood2014ethereum}. They possess three key characteristics: decentralization, immutability, and the ability to handle high financial values. Being decentralized ensures that no single entity controls them, promoting transparency and trust among diverse participants~\cite{atzei2017survey}. A smart contract is immutable, which means that it is extremely difficult and expensive to change its code once it has been deployed on the blockchain~\cite{wood2014ethereum}. A smart contract facilitates financial transactions, automates payment processing, enforces conditions for fund transfers, eliminates the need for intermediaries, and reduces transaction costs~\cite{destefanis2018smart}. These features make smart contracts valuable tools for creating secure, transparent, and efficient agreements within decentralized systems.

On Ethereum, smart contracts are written in Solidity~\cite{solidity-docs} language or Vyper language~\cite{vyperdocs}, which includes programmable functions and state variables similar to high-level programming languages. Smart contract code is used to force strict rules, commitments, and consequences for parties under several conditions, similar to traditional legal contracts. A smart contract uses parameters within the functions to carry out transactions effectively. For instance, in a smart contract designed to manage token transfers, such as transferring tokens from one user to another, the parameters for this transaction might encompass the sender's address, the recipient's address, and the specific amount of tokens intended for transfer. 

Based on the logic implementation of the smart contract, the status of variables changes during the execution of the function. In Ethereum, bytecode is generated from the Solidity contract code. The generated bytecode is then executed by the Ethereum Virtual Machine (EVM)~\cite{hirai2017defining, EthereumWhitePaper}. The generated bytecode is an assembly language that contains low-level instructions called opcodes~\footnote{https://ethereum.org/en/developers/docs/evm/opcodes/}. Eventually, each deployed contract is assigned a unique address on the blockchain network. As part of their verification and agreement of the new blocks, each node participating in the blockchain executes the bytecode of the contract.
Users invoke smart contracts deployed on Ethereum by sending a transaction that includes the contract's address, signature, and input parameters for the targeted function. This transaction is directed to the nearest Ethereum node for processing. Smart contracts can manifest as independent applications. For instance, smart contracts may serve as the core or the backend for DApps~\cite{metcalfe2020ethereum}.

\subsection{Inline Assembly in Smart contracts}
Inline assembly is low-level code that is allowed to be used in Solidity smart contracts~\cite{solidity-docs}. There are many reasons to use inline assembly in Solidity, including granting developers fine-grained control over their contracts by implementing libraries or performing operations not available in Solidity~\cite{solidity-docs}. Another important reason to use inline assembly with Solidity is to optimize the contract, especially when Solidity’s optimizer does not provide efficient code. Inline assembly is written in a language called Yul, which is a readable low-level language. Despite the additional functionalities that inline assembly allows developers to integrate into their contracts~\cite{solidity-docs}.
Recently, an empirical study by Chaliasos et al.~\cite{chaliasos2022study} showed that around 23\% of 49 million contracts contain inline assembly and it is used frequently to add functionality that is not available in Solidity and for gas optimization using specific code patterns. Notably, available static analysis tools that check Solidity contracts (e.g. Slither~\cite{Slither}, Smartcheck~\cite{Smartcheck}, Osiris~\cite{Osiris}, etc )only partially support inline assembly fragments.

\subsection{Smart Contract Security}
In spite of the benefits that smart contracts provide, they are vulnerable to various security threats~\cite{perez2022secure}. In the history of smart contracts, there have been numerous cyperattacks that lead to high financial losses, with large number of smart contracts falling victim to hacking incidents leading to an estimated \$6.45 billion in damages~\cite{chaliasos2024smart}. The first attack occurred in 2016 when an attack on DAO contracts led to the loss of over 3.6 million Ethers due to a re-entrancy vulnerability~\cite{DAOattack}. Due to the increasing frequency of attacks, smart contract security and trustworthiness have gained significant attention from scholars, leading to numerous studies on smart contract vulnerabilities~\cite{AtzeiSurvey}, smart contract analysis~\cite{Smartcheck}, and vulnerability detection methods~\cite{chen2020gaschecker}. 

\subsection{Large language models (LLMs)}
Transformers enabled the development of large language models such as BERT (Bidirectional Encoder Representations from Transformers)~\cite{devlin2018bert} and GPT (Generative Pre-trained Transformer)~\cite{Radford2018ImprovingLU}, which achieved remarkable performance across a wide range of NLP tasks, including text classification, sentiment analysis, language translation, and more. Based on the success of Transformers in NLP, scholars have expanded their application to programming languages. For instance, CodeBERT and CodeBERTA~\cite{feng2020codebert,codeberta} can capture semantic relationships between code and plain language. They offer generic code representations that enable various code automation tasks, providing developers with more resources for comprehending and producing code. BERT was pre-trained on a large corpus of English text, including Wikipedia. It was trained using two primary tasks: masked language modeling and next sentence prediction, where the model predicts if two given sentences follow each other in the text. Therefore, BERT understands the context of words deeply in sentences. DistilBERT is a distilled version of BERT~\cite{sanh2019distilbert}. The model's design involves the removal of token-type embeddings contributing to a more efficient and faster processing capability. DistilBERT also simplifies the pre-training process by eliminating the Next Sentence Prediction (NSP) mechanism, while retaining the essential Masked Language Modeling (MLM) mechanism.
GPT-4~\cite{openai2023gpt4}, developed by OpenAI, was trained on a diverse dataset that includes 8 million web pages in the English language. GPT-4 leverages advanced transformer architecture to understand and generate human-like text. T5~\cite{raffel2020exploring}, developed by Google, can perform various natural language processing tasks, such as text translation with high accuracy. By converting all tasks into a text-to-text format, T5 leverages a unified framework that simplifies training and improves performance across different NLP tasks. In summary, Transformer-based architectures, have enabled applications to analyze and detect code vulnerabilities using their deep contextual understanding and powerful NLP capabilities.

  \section{Method}
\label{sec:method}
\begin{figure*}[ht]
\centering
\includegraphics[width=.95\textwidth]{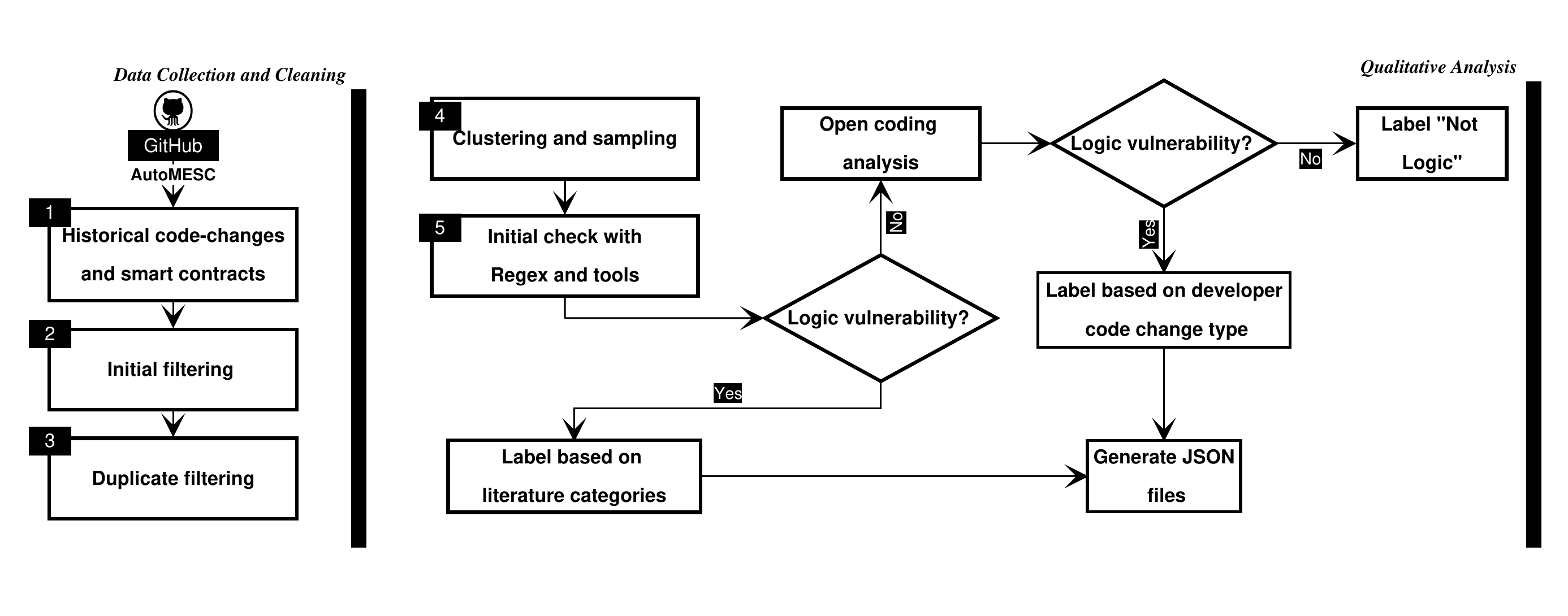}
\caption{Overview of Our Data Collection and Qualitative Analysis}
\label{fig:codepreprocessing}
\end{figure*}

  \begin{figure*}[ht]
\centering
\includegraphics[width=.95\textwidth]{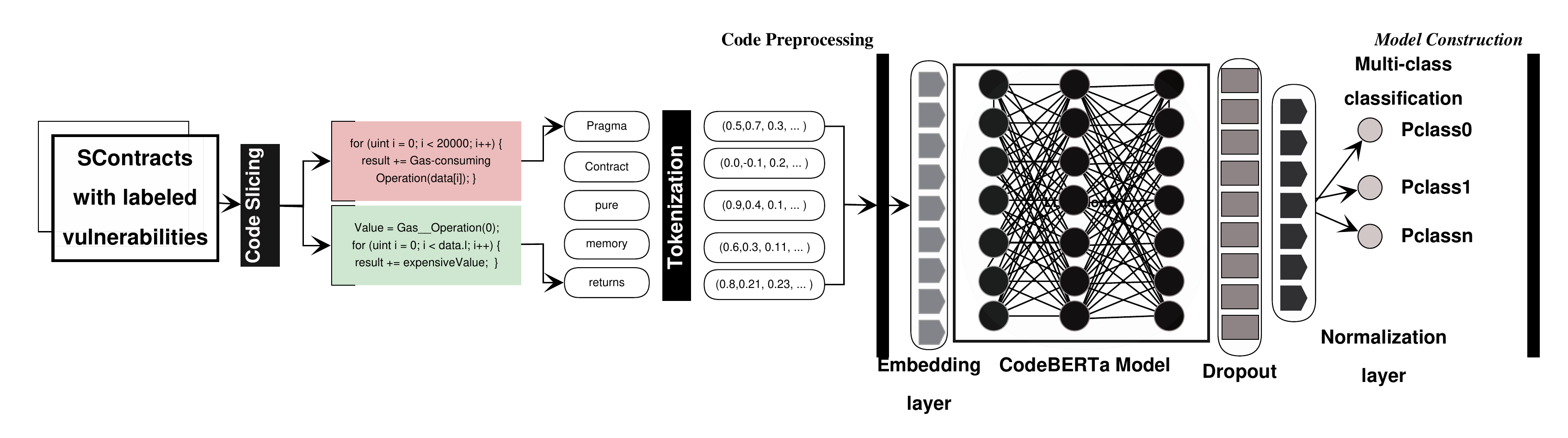}
\caption{Overview of Our Code Preprocessing and Model Construction}
\label{fig:codeberta}
\end{figure*}

In this section, we outline our research method. We begin by presenting the process of collecting and cleaning the smart contracts and code changes from GitHub. Afterward, we detail our qualitative approach to investigate the available logic vulnerabilities and corresponding mitigation strategies. 
Finally, we describe the code preprocessing, training, and evaluation phases for our proposed Sóley --- automated smart contract logic vulnerability detection.

\subsection{Data collection}
In this work, we collect an extensive dataset of real-life smart contract source code and code changes available on Open Source Software (OSS) projects hosted on GitHub using AutoMESC~\cite{soud2023automesc}. 

We selected GitHub, as it is the largest host of source code globally~\cite{wartschinski2022vudenc}. Moreover, smart contracts on GitHub are presented in their solidity source code form, which includes additional essential elements such as code changes, reviews, code vulnerabilities, commits, and fixes. These components significantly enhance the comprehensibility of the code during our analysis. GitHub as a version control system that is centered around commits. A commit that fixes a code vulnerability can be viewed as a fix, comprising two crucial pairs of source code: the initial version with code vulnerabilities and the subsequent updated and, ideally, fixed version as highlighted by Zhou et al. ~\cite{zhou2017automated}. Analyzing these commit differences (diffs) offers valuable insights into fixing the smart contract. 

Figure~\ref{fig:codepreprocessing}
shows our data collection process (left side). First, we collected contracts specifically written in Solidity (identified by the .sol extension) from various projects on GitHub
, including Ethereum official repository~\cite{ethereum-github} and Smatrbugs~\cite{smartbug}, resulting in a corpus of Solidity smart contracts. We also extracted code changes indicating code vulnerabilities or fixed code by gathering the commit differences (diffs) for each smart contract (.sol file), where available. However, some of these diffs were found only in comments and not in the actual contract code. We filtered out these comment-based changes, focusing solely on modifications made within the Solidity code. Our data collection and 
initial filtering were conducted through an automated pipeline utilizing AutoMESC~\cite{soud2022automesc}.

\subsection{Data cleaning and pre-processing}
We performed data cleaning by processing the collected data from the previous step to identify potentially duplicate smart contracts and code changes. In the duplicate filtering stage, where we retained contract URLs and their source codes, we proceeded to compare Solidity contract names and URIs. Through this process, we detected possible duplicates and removed them. 
Then we compiled the resulting contracts to remove the non-functional contracts using the Solidity compiler~\cite{solidity-docs}. Subsequently, we assigned unique hashes to distinct contracts and linked code change pairs with their corresponding contracts via AutoMESC. This approach yielded a corpus comprising a set of unique smart contracts, metadata related to these contracts, as well as the source code of the contracts and their associated code changes when available. The resulting dataset served as the input for our qualitative analysis.

\subsection{Quantitative Overview of the Collected Dataset}

Our dataset consists of smart contracts collected from GitHub, along with smart contracts with code changes obtained from GitHub code reviews and diffs. Table~\ref{tab:quantitative_data} presents the quantitative analysis of the collected smart contracts and code changes before and after filtering and duplicate removal. 

\begin{table}[h]
    \centering
    \small
    \begin{tabular}{l r}
        \toprule
        Description &  \\
        \midrule
        Total number of collected contracts & 54652 \\
          Total number of unique functional  contracts & 52458 \\
                    Total number of non-functional and duplicate  contracts & 2194 \\
        Total number of unique contracts with code changes & 6363 \\
        Total number of lines for all contracts with code changes & 1544321 \\
        Total number of vocabulary & 305377 \\
        Average number of lines & 242.703 \\
        Average number of vocabulary & 47.993 \\
        Min number of lines & 7 \\
        Max number of lines & 7829 \\
        Min number of vocabulary & 30 \\
        Max number of vocabulary & 24594 \\
        \bottomrule
    \end{tabular}
   \caption{Quantitative Overview of the Collected Dataset.}
    \label{tab:quantitative_data}
\end{table}

The total number of collected contracts written in Solidity is 54,652. Out of these, 2,194 contracts were non-functional or duplicates. As a result, we obtained 52,458 functional and unique contracts.
Among the smart contracts with at least one code change, there were a total of 13,000 contracts, including duplicates, and approximately 6,000 unique contracts. The total number of unique code changes is around 12,000. 

\subsection{Qualitative Analysis Approach}

Figure~\ref{fig:codepreprocessing} shows the steps we followed to perform our qualitative analysis. To analyze the logic vulnerabilities and their mitigation strategies in a large corpus of Solidity codes, we decided to cluster the code changes first based on general themes and then select random samples from each cluster for analysis, as we describe next.

We carefully examined a random set of 1000 unique code changes with varying vocabulary sizes and in various contracts in the first round to understand the types of changes developers typically make to their contracts. After that, we automatically clustered the code changes based on common characteristics (i.e., frequent keywords) into eight clusters using a Python script. The resulting clusters are described in Table~\ref{tab:code_clusters}.

\begin{table}[htbp]
\centering
\begin{tabular}{l r}
\toprule
\textbf{Cluster} & \textbf{Total} \\
\toprule
Standard Compliance & 1099 \\
Storage & 1005 \\
Pragma Changes & 306 \\
Contract Oracle & 147 \\
Assembly & 145 \\
Upgradability & 48 \\
Random & 1805 \\
Testing & 1445 \\
\toprule
\end{tabular}
\caption{Clustering of Smart Contract Code Changes}
\label{tab:code_clusters}
\end{table}

We then selected a random 10\% of code changes from each cluster, except for the upgradability and pragma clusters.
We selected all 48 code changes from the Upgradability cluster considering its small size. 
Additionally, we selected 48 code changes from the Pragma Changes cluster, after the initial analysis of the pragma cluster's code changes revealed no explicit logic vulnerabilities or mitigation strategies.

After sampling and clustering steps, we started the analysis of the code changes to obtain the types of logic vulnerabilities that exist in these changes, as shown in Figure~\ref{fig:codepreprocessing}. First, we checked if the available code vulnerabilities corresponded to any of the defined logic vulnerabilities in the literature, assigning labels accordingly. We extracted the defined logic vulnerabilities from \cite{sun2024gptscan,soud2022fly,zhang2020framework,zhang2023demystifying} and utilized static analysis tools such as Slither~\cite{Slither} and Smartcheck~\cite{Smartcheck} along with regular expressions to identify logic-related vulnerabilities. The generated labels were saved in a JSON file for each contract. Two reviewers (i.e., a doctoral student expert in smart contracts and software security, and a master's degree student expert in smart contracts and machine learning) examined the regular expressions in three rounds. If no logic vulnerability was identified in the code change, we employed an open coding method for analysis. The reviewers manually inspected the code changes, related contracts for context, and any developer comments. In cases where the code change belonged to a new logic vulnerability not identified by the literature,  we relied on developer comments or the type of code change itself to assign it a name. Otherwise, we labeled it as a non-logic-related vulnerability. The qualitative analysis resulted in labeled JSON files containing contracts, code changes, and their corresponding logic vulnerabilities, enabling us to extract categories of logic vulnerabilities and their mitigation strategies.

To support our automated logic vulnerability detection, a substantial amount of data is needed. To achieve this, we utilized regular expressions on the contracts we had, in addition to employing various analysis tools as we mentioned earlier to identify known logic vulnerabilities in smart contracts. If the tools' output aligns with the output from the regular expressions, we automatically labeled the vulnerability and generated JSON files. Each JSON file contains information about the detected lines with code vulnerabilities, including the vulnerability type, contract filename, and some metadata. In the manually labeled JSON file, we included an additional attribute that provides an explanation of why the detected issue is considered a code vulnerability. This explanation facilitates replication and verification by the research community.

\subsection{Code Pre-processing and Granularity Level Selection}
Solidity contracts, despite sharing identical tokens, may vary in cleanliness or susceptibility to code vulnerabilities. This highlights the need for a detailed analysis of individual lines of code, rather than relying solely on a top-down analysis of entire contracts. Moreover, previous research indicates that processing entire files for classification has limitations in effectively identifying code vulnerabilities~\cite{morrison2015challenges}. 

Therefore, our study adopts a fine granularity similar to Wartschinski et al.~\cite{wartschinski2022vudenc} by examining code tokens~\footnote{fundamental building blocks of the code, such as keywords, identifiers, operators, and literals.} responsible for logic vulnerabilities and their contextual usage by considering sequences of related code tokens. This approach enables us to precisely pinpoint the location of code vulnerabilities within sequential code, considering the interdependence of statements. Considering a specific token as a code vulnerability may result in many false positives. Instead, we assume that a token signifies a code vulnerability when utilized in a particular way. Therefore, we consider the surrounding tokens and lines, based on the lines we have in the generated JSON files. 

\subsection{Code Slicing and Tokenization}
After determining our study's granularity level, we proceeded with the code-slicing~\cite{weiser1984program} phase as shown in Figure~\ref{fig:codeberta}. This phase involved partitioning the code into two categories: negative slices containing code lines demonstrating logic vulnerabilities, and positive slices containing code lines without such vulnerabilities. We iterated through each contract and extracted the negative slices along with the corresponding type of logic vulnerability and the positive slices. 
The tokenizer utilized in our study is a Byte-level BPE tokenizer trained on the corpus using Hugging Face tokenizers~\footnote{https://huggingface.co/huggingface/CodeBERTa-small-v1}. 
Sóley operates on the full source code of the sliced smart contract, preserving the code exactly as it is. To transform a slice of code into a numerical representation, the tokenizer  first splits the code into a list of Solidity code tokens, including keywords (e.g., \textit{``contract"}, \textit{``function")}, identifiers (e.g., variable names), operators (e.g., \textit{``+"}, \textit{``-"}), and other language-specific constructs. The tokenizer provides a dedicated lexical scanner for various source codes, including Go, Python, and Java. For instance, it splits a line such as \textit{``function transfer(address \_to, uint256 \_value) public returns (bool success)"} into the tokens: \textit{``function"}, \textit{``transfer"}, \textit{``("}, \textit{``address"}, \textit{``\_to"}, \textit{``,"}, \textit{``uint256"}, \textit{``\_value"}, \textit{``)"}, \textit{``public"}, \textit{``returns"}, \textit{``("}, \textit{``bool"}, \textit{``success"}, \textit{``)"}, \textit{``;"}. Each of these tokens is then embedded, i.e., represented by a numeric vector. Consequently, a full section of code is transformed into a series of numbers.

\subsection{Model Selection and Construction}

Source code inherently consists of sequential data, where the effect of each statement depends on surrounding instructions. Hence, we need a model capable of learning features associated with code vulnerabilities from sequences of code tokens. Machine learning-based models, especially transformers, excel at representing the sequential nature of code while capturing its semantics. These models are adept at tasks similar to our task, as they can effectively model code and have been successfully utilized in similar contexts. Therefore, we selected CodeBERTa~\cite{codeberta} as the model for this work. The chosen model consists of 6 layers and 84M parameters. It is a RoBERTa-like transformer model that shares the same number of layers and heads as DistilBERT. The model was initialized from default settings and trained from scratch on the full corpus.

The selected CodeBERTa model architecture includes a series of layers and components designed to process input data effectively. The model begins with embedding layers, which map input tokens to high-dimensional representations. The first embedding layer has a vocabulary size of 52,000 and outputs 768-dimensional embeddings, while the second embedding layer has a vocabulary size of 514 and also outputs 768-dimensional embeddings.

Following the embedding layers, dropout layers are applied to prevent overfitting by randomly setting a fraction of input units to zero during training. Layer normalization layers are then applied to normalize the outputs of the previous layers, ensuring stable training.
The model also includes multiple convolutional layers, which perform convolution operations on the input embeddings. These convolutional layers are interleaved with dropout layers and layer normalization layers to improve the model's ability to capture complex patterns in the data.

An activation function called NewGELUActivation is applied after every second convolutional layer.  After the convolutional layers, additional dropout and layer normalization layers are applied to further regularize the model. Finally, it has a linear layer that maps the 768-dimensional input to a 6-dimensional output for multiclass classification.

\subsection{Model Training}
From the preprocessed data, we randomly selected 1000 samples of five logic vulnerabilities, all containing inline assembly fragments. However, for a sixth logic vulnerability, we could only select 592 samples because these were the only ones with inline assembly, while excluding those without.
To ensure consistency in the training process and assess the accuracy of our model and baseline models, we maintained identical hyperparameters throughout. The length of text sequences was limited to 512, and a batch size of 10 was employed. These parameters are interrelated, as they dictate the memory requirements for each batch. While a higher batch size can expedite training, it also necessitates more memory.
For optimization, we utilized the AdamW optimizer with a learning rate of 5e-5 and epsilon of 1e-8, as prescribed in the original GPT-2 paper. All models underwent 10 epochs of training, with evaluation statistics recorded at each iteration.
Training larger models locally presents challenges due to their substantial memory requirements. However, we mitigated this issue by utilizing smaller models with reduced batch sizes, enabling training on laptops with 16 gigabytes of RAM and 4 gigabytes of VRAM. Although training times were longer compared to GPU clusters, attempting to initialize larger models on GPUs risked memory constraints. While decreasing sequence length and batch sizes helped address this challenge, it also resulted in longer training times. For training, We split the data into a 75/25 ratio for training and evaluation, following the default setting of the train\_test\_split function from scikit\-learn.

 \section{Experiment Evaluation}
\label{sec:setup}
In this section, we describe our experiment and evaluation measures. Additionally, we present the selected baseline. 

\subsection{Metrics and Evaluation}
We trained each of the models on the same dataset, with the same settings and hyperparameters, and then ran an evaluation benchmark from the scikit-learn\footnote{\url{https://scikit-learn.org/stable/}} package. For the evaluation, we use the precision, recall, and F1 measure from the scikit-learn package, as well as an accuracy calculation. Accuracy was determined by accumulating correct predictions throughout the evaluation process and computing the average accuracy score at its conclusion.
\[
\text{Accuracy} = \frac{\text{Number of Correct Prediction}}{\text{Number of Samples}}
\]

The precision value describes the models ability to not falsely label negative samples as positive. Recall, on the other hand, measures the model's ability to accurately capture all positive samples.

\[
\text{Precision} = \frac{\text{True Positives}}{\text{True Positives} + \text{False Positives}}
\]

\[
\text{Recall} = \frac{\text{True Positives}}{\text{True Positives} + \text{False Negatives}}
\]

Finally, F1-measure represents the harmonic mean of the precision and recall. This is a type of average, but for each of the samples instead of counting them together.
\[
\text{F1} = \frac{2 * \text{True Positives}}{2 * \text{True Positives} + \text{False Positives} + \text{False Negatives}}
\]

The accuracy in this case reflects the overall performance over all the samples, while precision, recall, and F1-measure were evaluated for each type of logic vulnerability separately. This approach highlights whether the model struggles in detecting specific types of logic vulnerabilities.

\subsection{Baseline Selection}
\textbf{Baseline 1:} We employ the base model used by Sun et al.\cite{sun2023assbert} for the task of detecting smart contract vulnerabilities. The selected model leverages various implementations of the BERT model. This model has been widely recognized as a ubiquitous baseline in NLP experiments for its widespread use and robust performance, as noted by Rogers et al.~\cite{rogers2020primer}. We adopt this baseline model for its application in detecting vulnerabilities in smart contracts. Following the description in the paper, we utilized the base BERT architecture since the original code was not provided. 

\section{Results}
\label{sec:results}
This section presents our research findings and addresses the proposed research questions. We begin by analyzing logic-related vulnerabilities in real-world smart contract code changes. Next, we evaluate Sóley's automated detection capabilities across various vulnerability types and compare the results with other well-known LLMs performing the same task. Finally,  we summarize the strategies developers employ to mitigate potential logic vulnerabilities.

\subsection{Answers to (RQ1): Logic Vulnerability Types}
To address the RQ1, regarding logic vulnerabilities in our dataset, we begin by conducting a qualitative analysis of the logic-related vulnerabilities identified by the selected tools in our dataset. Overall, the total number of vulnerabilities in our dataset is 428,568. Of these, 171,180 are logic-related code vulnerabilities.

We present a comprehensive analysis of the vulnerabilities available in our dataset in the Appendix.
Before starting our manual analysis, we reviewed publications on logic vulnerabilities in smart contracts to identify known vulnerabilities, as summarized in Table~\ref{tab:related_papers}.
We identified 17 logic vulnerabilities from the literature. We present the list of identified logic vulnerabilities along with their definitions and corresponding references in Table~\ref{tab:vulnerabilities_logic}.

\begin{table*}[ht]
\begin{tabular}{lllllll}
\hline
\textbf{Papers}                                                          & \multicolumn{3}{l}{\textbf{Venue}} & \multicolumn{3}{l}{\textbf{Logic Vulnerabilities}}                                                               \\ \hline
Sun et al.~\cite{sun2024gptscan}          & \multicolumn{3}{l}{ICSE}           & \multicolumn{3}{l}{Defined 9 scenarios of known logic vulnerabilities in smart contracts.}                        \\
Soud et al.~\cite{soud2024fly}            & \multicolumn{3}{l}{EMSE}           & \multicolumn{3}{l}{Identified 4 logic vulnerabilities in smart contracts.}                                        \\
Chaliasos et al.~\cite{chaliasos2024smart}&
\multicolumn{3}{l}{ICSE}           & \multicolumn{3}{l}{Tested existing smart contract tools for logic vulnerabilities, errors, and sanity checks. }    \\

Zhang et al.~\cite{zhang2020framework}    & \multicolumn{3}{l}{ICSME}          & \multicolumn{3}{l}{Identified 8 logic vulnerabilities in smart contracts and categorized them into 4 categories.} \\ \hline
\end{tabular}
\caption{Publications Discussing or Identifying Logic Vulnerabilities in Smart Contracts}
\label{tab:related_papers}
\end{table*}
We then proceeded to manually label 614 code changes sampled from the clusters we have as mentioned in the method section~\ref{sec:method}, from which we were able to define nine logic vulnerabilities. In the following section, we explain each defined logic vulnerability. While certain vulnerabilities may manifest in various forms within smart contract code, due to space constraints in this paper, we present code examples when needed. Additional examples and various forms of these vulnerabilities can be found in our dataset, where the full contracts and code changes are available.

\subsubsection{Bypassing Solidity access control via inline assembly}
Inline assembly in Solidity allows developers to directly integrate assembly code within their Solidity contracts to effectively execute certain operations with less gas cost and added flexibility~\cite{chaliasos2022study} (\textit{see} the Background section~\ref{sec:preliminaries}). However, one significant risk is that inline assembly can bypass Solidity's access control mechanisms, affecting the logic and flow of the contract. Solidity's access control mechanisms typically depend on modifiers or conditional statements in Solidity code to enforce restrictions on function calls based on certain conditions~\cite{solidity-docs}. Attackers or unauthorized users could potentially use inline assembly logic to gain unauthorized access to restricted function logic, even if they do not meet the specified conditions for access according to the Solidity code. In the following, we illustrate this vulnerability with example listing~\ref{List:Inlineassembly}. In the example, we have a \textit{``setSecret"} function in Solidity that utilizes an access control mechanism, specifically employing a \textit{``require"} statement to ensure that only admin users can set a secret. No other user should have the privilege to set the secret, making the admin the sole entity with this capability. However, we also have a \textit{``setSecretWithAssembly"} function that bypasses the implemented access control mechanism. This is achieved by directly accessing the \textit{``isAdmin"} mapping slot using inline assembly. Consequently, an unauthorized user can call \textit{``setSecretWithAssembly"} and set the secret without being checked against the admin role. This leads to the unintended usage of the logic implemented in the contract and eventually compromise its security.

\begin{lstlisting}[language=Go,  label={List:Inlineassembly}, caption= Bypassing Solidity access control via inline assembly]

  function setSecret(uint256 _secret) public {
        require(isAdmin[msg.sender], "Only admin can set secret");
        
        // Some code for setting secret here
    }
function setSecretWithAssembly(uint256 _secret) public {
    assembly {
        // Load the isAdmin mapping slot
        let slot := sload(isAdmin.slot)

        // Logic vulnerability 
        //Load the value from the mapping using slot and msg.sender
        let isAdminValue := sload(add(slot, mul(and(isZero(shr(96, calldataload(0))), 0xffffffff), 0x20)))  }}}
\end{lstlisting}
\begin{table*}[h!]
\centering
\small
\begin{tabular}{>{\raggedright\arraybackslash}m{4.5cm}|>{\raggedright\arraybackslash}m{10cm}|>{\raggedright\arraybackslash}m{1cm}}
\toprule
\textbf{Vulnerability} & \textbf{Description}& \textbf{Ref.} \\
\toprule
Unauthorized Transfer & A smart contract allows assets to be transferred without proper authorization, leading to potential theft or loss of funds. &~\cite{sun2024gptscan} \\
\hline
Risky First Deposit & The first deposit into a contract sets critical parameters that can be exploited by the first user, creating unfair advantages or security risks. & ~\cite{sun2024gptscan} \\
\hline
Price Manipulation  & Attackers manipulate asset prices by exploiting the liquidity pool algorithms in decentralized exchanges using Automated Market Makers (AMMs). & ~\cite{sun2024gptscan} \\
\hline
Front Running & An attacker observes a pending transaction and places a similar transaction with higher fees to be processed first, leading to financial gain at the expense of the original transaction. & ~\cite{sun2024gptscan} \\
\hline
Wrong Interest Rate Order & Interest rates are calculated or applied in an incorrect sequence, leading to misallocation of interest payments or incorrect financial incentives. & ~\cite{sun2024gptscan} \\
\hline
Wrong Checkpoint Order & Errors in the sequence of recorded checkpoints within a contract, leading to incorrect contract behavior or mismanagement of funds. & ~\cite{sun2024gptscan} \\
\hline
Slippage & The difference between the expected price of a trade and the actual price executed, often causing losses for traders due to low liquidity or rapid market movements. & ~\cite{sun2024gptscan} \\
\hline
Greedy Contract & A smart contract that consumes excessive resources (e.g., gas) during function calls, potentially depleting user funds. & ~\cite{soud2024fly}  \\
\hline
DoS by External Contract & A smart contract relies on an external contract that can be manipulated or fails to respond, causing a denial of service. & ~\cite{soud2024fly}  \\
\hline
Call to the Unknown & A smart contract calls a function on an unknown or user-specified address without proper validation, potentially executing malicious code. & ~\cite{soud2024fly},~\cite{zhang2020framework}  \\
\hline
Transaction Order Dependency (TOD) & The outcome of a contract depends on the order of transactions within a block, allowing attackers to gain an unfair advantage by influencing transaction order. & ~\cite{soud2024fly},~\cite{zhang2020framework}  \\
\hline
Returning Results Using Assembly in the Constructor & Using low-level assembly language in a constructor to return data, bypassing typical restrictions and potentially leading to complex, less readable code. & ~\cite{zhang2020framework}\\
\hline
Specify Function Variable as Any Type & Using a generic or ambiguous data type for function parameters, leading to unexpected behaviors or security vulnerabilities. & ~\cite{zhang2020framework}\\
\hline
DoS by Gas Limit & Exploiting the gas limit by creating situations where the required gas exceeds the block gas limit, causing transactions to fail. &~\cite{zhang2020framework}\\
\hline
DoS by Complex Feedback Function & Invoking a function involving complex computations or recursive calls that consume excessive gas, preventing the function from completing. & ~\cite{zhang2020framework}\\
\hline
DoS by Non-Existent Function & Calling a function that does not exist, causing the transaction to fail and revert, leading to a denial of service. & ~\cite{zhang2020framework}\\
\hline
Storage Overlap Attack & Different variables or mappings share the same storage slot due to incorrect definitions, allowing manipulation of one variable by changing another.&  ~\cite{zhang2020framework}\\
\toprule
\end{tabular}
\caption{Identified logic vulnerabilities in the literature, along with their definitions and corresponding references.}
\label{tab:vulnerabilities_logic}
\end{table*}

\subsubsection{State manipulation via inline assembly:}
This vulnerability is about manipulating state within inline assembly, particularly under certain conditions that may not be immediately apparent. Such actions can lead to state inconsistencies within the logic of the contracts. These inconsistencies introduce uncertainty into the contract functionality and logic flaws. Consequently, developers and auditors may face difficulties in fully assessing the contract's logic and behavior. In listing~\ref{List:statemipulation}, The \textit{``conditionalManipulation"} function employs inline assembly to modify the \textit{``manipulatedValue"} only if the input \_value exceeds a specific threshold, in this instance, 1000. This manipulation has the potential to disrupt the logic of the \textit{``withdraw"} function, which verifies \textit{``manipulatedValue"} before permitting withdrawals. If \textit{``manipulatedValue"} is altered to an unexpected value, the \textit{``withdraw"} function reverts, which shows how state manipulation can alter the logic of the contract.
\begin{lstlisting}[language=Go,  label={List:statemipulation}, caption= Incorrect Validation of Oracle Data]
function conditionalManipulation(uint256 _value) public {
    assembly {
        // Conditionally store value in manipulatedValue's slot
        if gt(_value, 1000) {
            sstore(manipulatedValue.slot, _value)}    }
    emit Manipulation(_value);}

function withdraw(uint256 _amount) public {
    require(balances[msg.sender] >= _amount, "Insufficient balance");
    require(totalSupply >= _amount, "Insufficient total supply");

    // Check if manipulatedValue has been set to an unexpected value
    if (manipulatedValue > 1000) {
        revert("State manipulated, cannot proceed");}
    balances[msg.sender] -= _amount;
    totalSupply -= _amount;
    payable(msg.sender).transfer(_amount);
    emit Withdraw(msg.sender, _amount);}

\end{lstlisting}

\subsubsection{Incomplete Standard function implementation }
This vulnerability occurs when a smart contract fails to implement all the required functions of a standard or modifies them~\cite{ERC2024}, resulting in an incomplete adherence to the standard. As a consequence, the contract's logic is compromised, potentially leading to unexpected behavior when interacting with other contracts that assume full compliance with the standard.

Consider an ERC-20~\cite{ERC2024} token contract that does not properly implement the "transferFrom" function, leading to a vulnerability that can cause unexpected behavior when transferring tokens between addresses. Therefore, it can potentially compromise the logic of other smart contracts or dApps built on top of the ERC-20 token standard. In this case, the logic of the ERC-20 token contract itself is compromised due to the improper implementation of the "transferFrom" function. For instance, if the "transferFrom" function fails to properly deduct the transferred token amount from the sender's balance or update the recipient's balance, it can lead to inconsistencies in token balances and potentially allow unauthorized token transfers. As a result, the contract's core logic for managing token transfers and maintaining token balances may become unreliable, impacting the overall functionality and security of the token contract.

\subsubsection{Incorrect validation of oracle data}
In blockchain, oracles serve as crucial links between off-chain data sources and on-chain smart contracts~\cite{solidity-docs}. They facilitate the flow of essential data inputs to smart contracts, enabling them to execute predefined actions based on real-world events.
 However, when the validation process for oracle data is flawed or when a contract overly relies on a single oracle, it introduces vulnerabilities into the contract's logic. For instance, if a contract fails to adequately validate data obtained from an oracle, it may accept inaccurate or malicious inputs, leading to incorrect contract execution. Similarly, depending solely on one oracle introduces a single point of failure, making the contract vulnerable to manipulation. In both cases, the contract's logic becomes compromised, as it cannot ensure the integrity of the data it relies upon for decision-making. In listing~\ref{List:Incorrectvalidation}, the \textit{``updatePrice"} function does not validate the \textit{``newPrice"} received from the oracle. As a result, any erroneous or malicious data provided by the oracle will be accepted by the contract logic and stored without verification. This oversight can result in unexpected logic outputs or other unintended behaviors, particularly if the contract relies on the price data for critical decision-making processes.
\begin{lstlisting}[language=Go,  label={List:Incorrectvalidation}, caption= Incorrect Validation of Oracle Data]
interface Oracle {
    function getPrice() external view returns (uint256);}
contract PriceConsumer {
    address public oracle;
    uint256 public price;
    
    event PriceUpdated(uint256 newPrice);
    constructor(address _oracle) {
        oracle = _oracle;}
    
    function updatePrice() public {
        // Fetch price from the oracle
        uint256 newPrice = Oracle(oracle).getPrice();
        // No validation on the price data received from the oracle
        price = newPrice;
        emit PriceUpdated(newPrice); }
    function getPrice() public view returns (uint256) {
        return price;} }

\end{lstlisting}
\subsubsection{Stale oracle data}
This vulnerability arises when a smart contract relies on outdated data from oracles, leading to logical decisions based on obsolete information. For instance, trading contracts dependent on price feed oracles may execute transactions using outdated prices, resulting in unwanted logic. Moreover, inadequate handling of unexpected or malformed responses from oracles can cause unintended logic. For example, if an oracle returns a negative or zero value unexpectedly, it may disrupt the contract's behavior. Therefore, a contract logic should gracefully handle such scenarios is crucial for maintaining logic integrity and preventing undesirable outcomes.

\subsubsection{Enumerating storage slot logic}
The logic of enumerating slots in the contract storage is significant. Developers can enumerate their contract storage slots using a pattern that is well-known and predictable. Therefore, attackers can access or guess the storage content and manipulate it in their favor. This manipulation can disrupt the intended flow of the contract logic by altering critical data stored in specific storage slots. For instance, using common offsets for storing data crucial for the contract logic enables attackers to target specific storage slots and thus execute the logic as they wish.

\subsubsection{Insecure upgrade authorization logic}

This vulnerability concerns the logic managing the authorization of contract upgrades, allowing unauthorized parties to approve or trigger upgrades. It may arise from weak or predictable authorization mechanisms, easily bypassed by users or unauthorized entities. Consequently, this can lead to unauthorized modifications, manipulation of contract logic, or even complete contract takeover by malicious actors.

\subsubsection{Inconsistent state transition checks}
This logic-related vulnerability can occur during contract upgrades. It involves inadequate checks for the contract's current state or version, which can lead to unexpected behaviors in the execution of the contract logic. For instance, if the contract fails to properly verify its current state or version before initiating an upgrade, it may result in inconsistencies or vulnerabilities in the contract's logic. As a consequence, the contract may experience malfunction, data corruption, or unexpected behaviors due to incorrect state transition checks during the upgrade process.

\subsubsection{Insecure rollback mechanisms}
This vulnerability occurs during the contract upgrading process, particularly when the implemented rollback mechanisms are inadequate or insecure. This can result in vulnerabilities or unexpected logical execution if upgrade failures or rollbacks occur. Rollback mechanisms in smart contracts typically refer to the ability to revert changes made to the contract state in case of errors or failures. For example, an incorrect implementation of robust rollback mechanisms may fail to properly revert state changes and ensure contract logic integrity in the event of upgrade failures or rollbacks, potentially leaving the contract in an inconsistent state. The impact of this vulnerability can manifest as contract malfunction due to inadequate rollback mechanisms that fail to properly handle upgrade failures or rollbacks.

\begin{tcolorbox}[colback=gray!10!white,colframe=gray!60!black,title=Summary] 
\begin{itemize}
    \item  Code changes provide valuable insights into logic vulnerabilities, evident from detecting numerous logic-related vulnerabilities. However, given the intricate nature of these vulnerabilities, it is essential to analyze code changes alongside the contract's source code and related contracts to gain a comprehensive understanding.
\item  Manual analysis of code changes, particularly when examining the business logic, proved challenging. Nonetheless, it revealed novel logic vulnerabilities, highlighting the depth of vulnerabilities that automated methods may overlook.
\item Logic vulnerabilities in smart contracts range from syntactically detectable vulnerabilities, easily identified by tools or using regular expressions, to semantic-specific vulnerabilities that demand a profound understanding of the semantic of the contract and its business logic and context.

\end{itemize}
\end{tcolorbox}

\subsection{Answers to (RQ2): Automated detection of logic-related vulnerabilities}
To answer RQ2, we utilized labeled vulnerabilities as described in the method section~\ref{sec:method}. We split the data into a 75/25 ratio for training and evaluation, following the default setting of the train\_test\_split function from scikit\-learn. The training method followed is outlined in Section~\ref{sec:method}. During the training process, we evaluated our approach and related models on an unseen test set to assess their accuracy in detecting various types of logic and general vulnerabilities within the contract code.

\begin{table}[h!]
\centering
\small
\begin{tabular}{lccc}
\toprule
                    & Precision & Recall & F1 \\
\midrule
Reentrancy (RE)         & 0.89      & 0.88   & 0.89     \\
Uninitialized Local Variables (UL)       & 0.93      & 0.93   & 0.93     \\
Recursive Calls in Loops (CLP)       & 0.92      & 0.95   & 0.94     \\
Incorrect Low-Level Calls (LLC)& 0.94      & 0.93   & 0.94     \\
Locked Ether (LE)       & 0.97      & 0.91   & 0.94     \\
Incorrect Equality  Check (IE)    & 0.81      & 0.88   & 0.84     \\
\bottomrule
\end{tabular}
\caption{Sóley Performance Metrics for Different Vulnerability Categories}
\label{tab:performance_metrics}
\end{table} 
To effectively detect logic vulnerabilities within smart contracts, LLMs require a substantial number of labeled instances comprising numerous labeled instances of such vulnerabilities. Instead of a binary classification approach (vulnerable or not), we need multi-class classification to pinpoint the specific type of vulnerability present, if any. We have curated a selection of vulnerabilities, all of which include inline assembly fragments and are categorized as logic-related by~\cite{zhang2020framework}, which demand nuanced comprehension of context and semantics. These include Locked Ether (LE) and Incorrect Equality Check (IE). In contrast, we have included three vulnerabilities easily detectable via pattern recognition. In addition, we have included Reentrancy (RE), as a recent empirical analysis by Chalias et al.~\cite{chaliasos2024smart} highlighted that all preventable attacks on smart contracts were related to RE vulnerabilities. Moreover, we have omitted business logic-related vulnerabilities due to labeling challenges and the unique nature of each contract's logic.  Our primary objective is to detect well-known logic vulnerabilities dependent on contextual understanding and compare the performance of LLMs in detecting them against vulnerabilities detectable through static analysis.

\begin{figure}[ht] 
\centering
\includegraphics[width=0.35\textwidth]{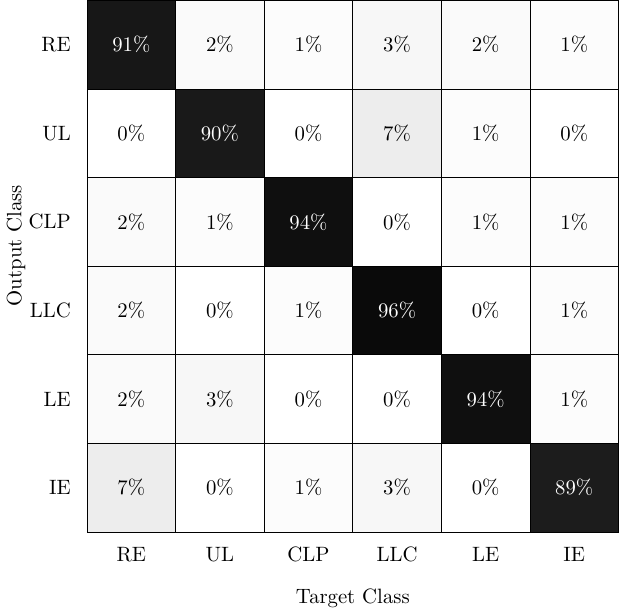}
\caption{Confusion matrix for the classification results of Sóley. Note: RE refers to Reentrancy, UL to Uninitialized Local Variables, CLP to Recursive Calls in Loops, LLC to Incorrect Low-Level Calls, LE to Locked Ether, and IE to Incorrect Equality Check.}
       \label{fig:Confusion1}
\end{figure}
The results in Table~\ref{tab:performance_metrics} show that Sóley excels in detecting Uninitialized Local Variables (UL), Recursive Calls in Loops (CLP), and Incorrect Low-Level-Calls (LLC). Precision and recall values consistently above 0.90, resulting in F1-measure of 0.93 and above. This reveals a balanced performance, effectively identifying true positives while minimizing false positives and negatives. These vulnerabilities are characterized by recognizable patterns or behaviors. However, vulnerabilities such as LE and IE show slightly lower performance measures, indicating Sóley's challenge in accurately identifying instances of these vulnerabilities. For instance, LE vulnerabilities involve nuanced contextual considerations regarding fund locking and withdrawal conditions within smart contracts. Similarly, detecting IE vulnerabilities may require a deeper understanding of how the contract handles money or value comparisons. For RE vulnerabilities, although the precision and recall are slightly lower, Sóley still achieves a high F1-measure of 0.89. We believe this is due to the potential variation in the entry points based on the contract logic. Nonetheless, Sóley demonstrates strong overall effectiveness in detecting vulnerabilities within smart contracts.

In Figure~\ref{fig:Confusion1}, we present the confusion matrix of Sóley classification results. We analyze the misclassification (i.e., false positives) to determine which vulnerabilities are being misclassified by Sóley. Notably, Sóley misclassifies IE as RE in 7\% of cases. This suggests that many instances labeled as IE are instead identified as RE. 
\begin{table}[ht]
\centering
\small
\begin{tabular}{llllllll}
 \toprule
                                                               & \multicolumn{7}{c}{\textit{Epoch 10}}                                                                                                                                                                                                                                          \\
                                                               & {\color[HTML]{44546A} \textbf{RE}} & {\color[HTML]{44546A} \textbf{UL}} & {\color[HTML]{44546A} \textbf{CLP}} & {\color[HTML]{44546A} \textbf{LLC}} & {\color[HTML]{44546A} \textbf{LE}} & {\color[HTML]{44546A} \textbf{IE}} & {\color[HTML]{44546A} \textbf{Acc.}} \\
\midrule
\\
{\color[HTML]{808080} \textit{Baseline}}  & 0.78                               & 0.84                                & 0.87                                & 0.91                                  & 0.89                                & 0.66                                & 0.84                                     \\
{\color[HTML]{808080} \textit{GPT2}}          & 0.85                               & 0.88                                & 0.88                                & 0.93                                  & 0.92                                & 0.81                                & 0.88                                     \\
{\color[HTML]{808080} \textit{T5-Base}}              & 0.85                               & 0.86                                & 0.88                                & 0.94                                  & 0.89                                & 0.78                                & 0.87                                     \\
{\color[HTML]{808080} \textit{Sóley}} & \textbf{0.9}                       & \textbf{0.91}                                & \textbf{0.93   }                             & \textbf{0.95}                         & \textbf{0.95}                       & \textbf{0.9}                        & \textbf{0.93}                            \\
{\color[HTML]{808080} \textit{DBERT}}          & 0.83                               & 0.88                                & 0.86                                & 0.93                                  & 0.9                                 & 0.8                                 & 0.87   \\                            \bottomrule   
\end{tabular}
\caption{Model evaluation in terms of F1-measure and accuracy (Acc.). Note: DBERT refers to DistilBERT. RE stands for Reentrancy, UL for Uninitialized Local Variables, CLP for Recursive Calls in Loops, LLC for Incorrect Low-Level Calls, LE for Locked Ether, and IE for IE Check.}
\label{tab:models_performance_FAtable}
\end{table}
The resulting misclassification between these two vulnerabilities can be due to several reasons. Firstly, IE and RE vulnerabilities might share similar patterns, making it difficult for the model to distinguish between them. For example, both vulnerabilities can involve complex control or data flow, which may confuse the model. Secondly, the training data might not adequately represent the differences between these vulnerabilities, causing the model to generalize poorly to new instances.  Lastly, the simplicity of Sóley model architecture may prevent the model from capturing nuanced differences between these two vulnerability types. However, since the percentage of misclassification is relatively small, we believe that there are patterns in these two vulnerabilities that share similarities in terms of data flow and the training data does not provide the model with clearly distinct features.
\begin{figure*}
	\centering
     \subfloat[][RE]
	{{\includegraphics[width=50mm]{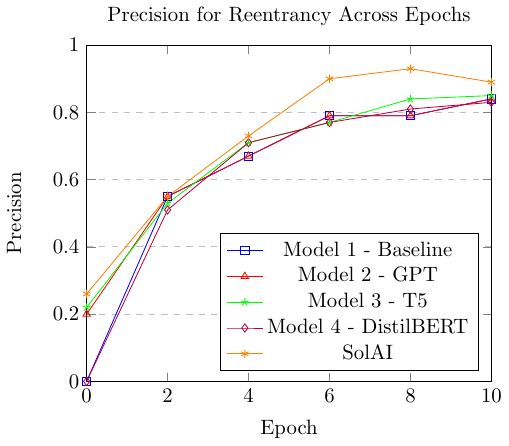}}}
      \subfloat[][UL]	{{\includegraphics[width=50mm]{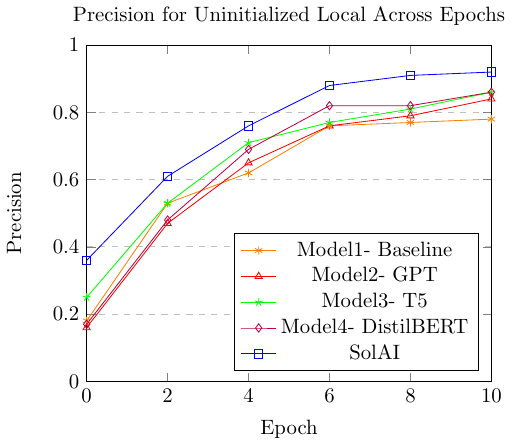}}}
        \subfloat[][CLP]
	{{\includegraphics[width=50mm]{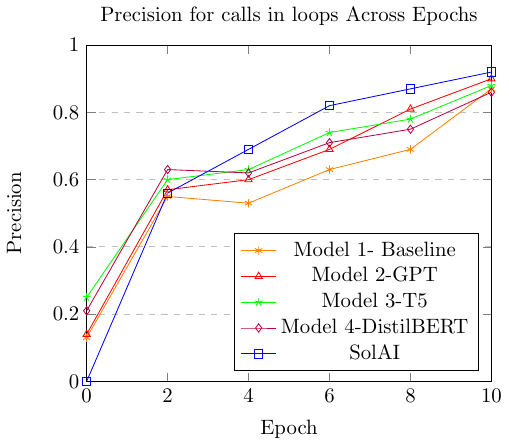}}}

      \subfloat[LE]
 {{\includegraphics[width=50mm]{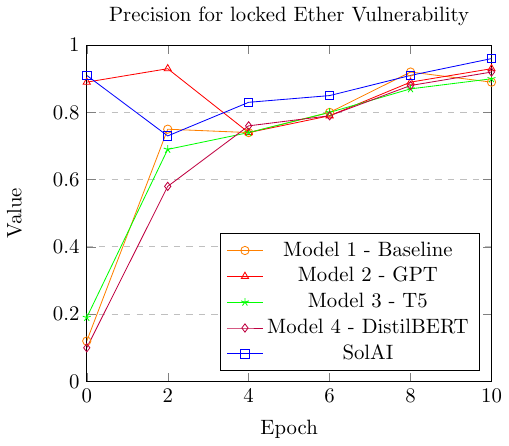}}}
      \subfloat[][IE]
 {{\includegraphics[width=50mm]{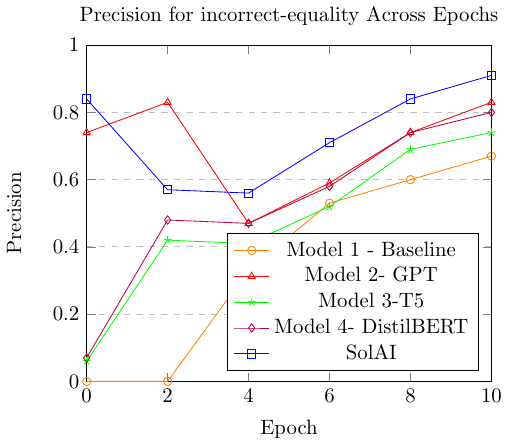}}}
       \subfloat[][LLC] 
 {{\includegraphics[width=50mm]{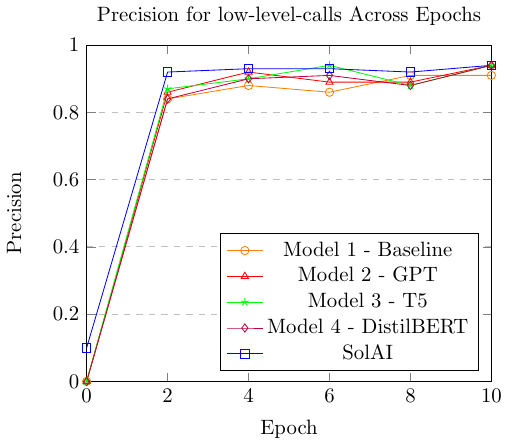}}} 
\caption{Precision rates across epochs for selected vulnerabilities.}
\label{fig:performance_across_epochesP}
\end{figure*}
\begin{figure*}
	\centering
     \subfloat[][RE]
	{{\includegraphics[width=50mm]{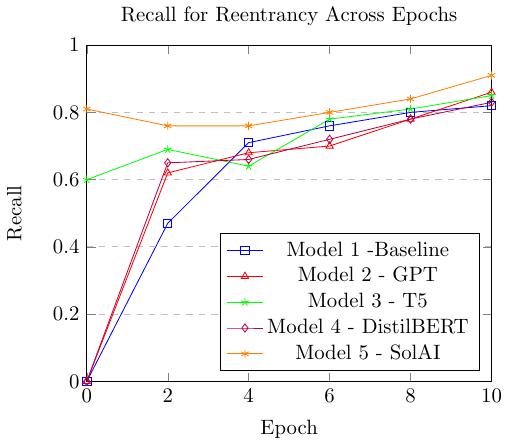}}}
      \subfloat[][UL]	{{\includegraphics[width=50mm]{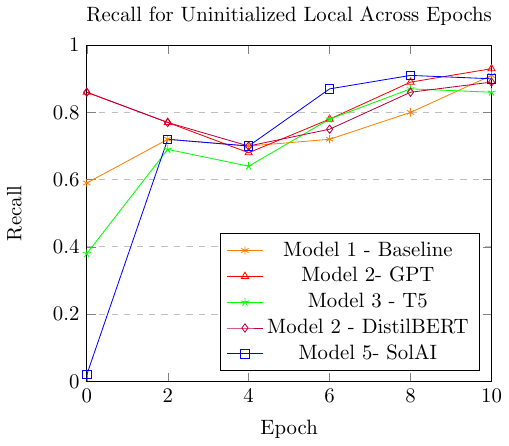}}}
        \subfloat[][CLP]
	{{\includegraphics[width=50mm]{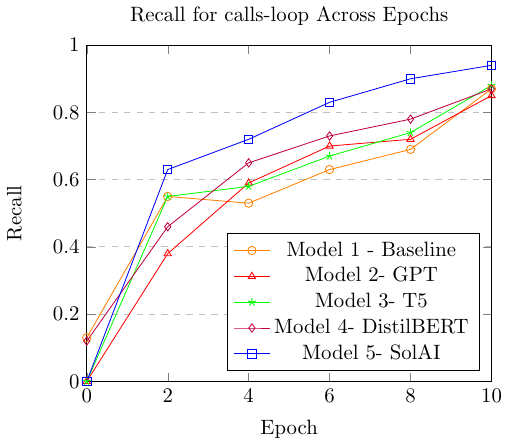}}}

      \subfloat[LLC]
 {{\includegraphics[width=50mm]{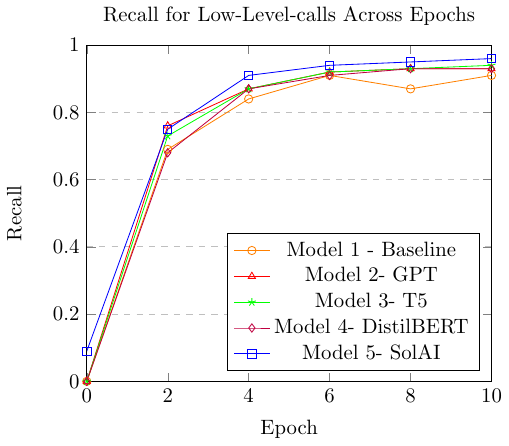}}}
      \subfloat[][IE]
 {{\includegraphics[width=50mm]{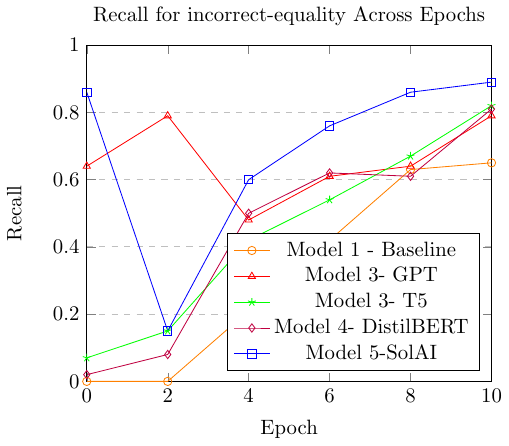}}}
       \subfloat[][LE] 
 {{\includegraphics[width=50mm]{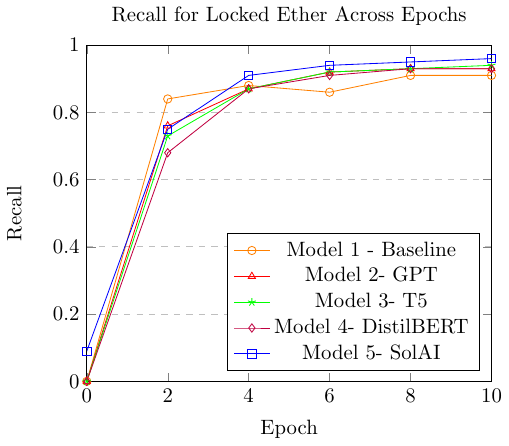}}} 
\caption{Recall rates across epochs for selected vulnerabilities.}
\label{fig:performance_across_epochesR}
\end{figure*}

Additionally, we observe a 3\% misclassification rate from low-level calls to uninitialized local vulnerabilities. The misclassification between these two types can be explained by instances where low-level function calls may share similarities with scenarios where variables are used without being properly initialized, such as both include accessing memory addresses, leading to confusion. 

To determine how well Sóley performs in comparison to the baseline and other related models, 
we fine-tuned the selected models for 10 epochs to maintain fair comparison. We also used the same hyperparameters for all models and the same training evaluation percentages. Table~\ref{tab:models_performance_FAtable} compares the selected models in terms of F1-measure along with the accuracy at epoch number 10. Our results indicate that Sóley outperforms the selected models and baseline in terms of accuracy, with a percentage of 5\%-9\% of improvement. We believe that is because our selected model was pre-trained on different programming languages, including object-oriented languages, that share similarities to Solidity syntax~\cite{atzei2017survey}. 

Regarding the F1-measure, we observe that all models struggle with the RE, it maybe as a result of determining the entry point in the logic of the contracts. So mostly the model identifies functions with the contract as susceptible to RE when they are not. Most of the models also face false positives in the IE, and that is because of erroneous identification of valid equality comparisons as vulnerabilities. The low-level calls vulnerability has the highest F1 measures across all models, this vulnerability arises when low-level function calls are made without proper validation, leading to security risks such as buffer overflows or injection attacks. It seems that all models recognize the LLC vulnerabilities' patterns, which are the easiest to detect. Finally, we can see that all models share a close trend regarding the detection performance for all vulnerabilities at epoch 10 in terms of F1-measures.

To study the performance of the selected models across different epochs we tracked the precision for all models per vulnerability as shown in Figure~\ref{fig:performance_across_epochesP}. The first six charts provide insights into the precision of various models across epochs for detecting different vulnerabilities. Across all vulnerabilities, we observe an overall increasing trend in precision as the training epochs progress. Therefore, while there may be variability in performance across different vulnerabilities, the models consistently improve in their ability to detect vulnerabilities over time, with some vulnerabilities being easier to detect than others. Similarly, recall rates in Figure~\ref{fig:performance_across_epochesR} reveal a notable increase in the performance of the selected models. However, for the incorrect-equality vulnerability, we see an unusual pattern where Sóley's recall starts high at 0.86, drops significantly at epoch 2, and then gradually recovers, indicating potential instability or sensitivity to early training stages. In contrast, DistilBERT exhibits a much lower starting recall for incorrect-equality but demonstrates a steady improvement, highlighting its robustness in learning over time despite initial setbacks. T5 also shows a unique pattern with a slow start in recall for incorrect-equality, but catches up significantly by epoch 10. 
However, with our dataset, the models began to overfit after 10 epochs, leading us to stop at this point. We believe there is potential for further accuracy improvements with extended training and larger datasets, especially for large models such as T5 and GPT.

\begin{tcolorbox}[colback=gray!10!white,colframe=gray!60!black,title=Summary]
\begin{itemize}
\item Our Sóley automated detection for logic-related vulnerabilities outperforms the baseline and the state-of-the-art related LLMs in terms of accuracy, with a percentage of 5\%-9\% of improvement.
\item Our experiments demonstrate that LLMs can effectively detect several types of vulnerabilities without excessive feature engineering. However, all models struggle with detecting RE vulnerabilities, most likely due to insufficient context or examples in the training set.
\end{itemize}
\end{tcolorbox}

\subsection{Answers to (RQ3): Mitigation strategies in smart contracts}
To answer this RQ, we comprehensively summarize common strategies followed by developers to address potential logic vulnerabilities. In the following, we provide detailed descriptions of each mitigation strategy. Moreover, many examples can be found in our dataset with the labeled code changes.

\textbf{S1: Optimizing and Simplifying Bitwise Operations in Inline Assembly.}
This strategy involves reviewing and replacing unnecessary bitwise shift operations within inline assembly with built-in Solidity operators. Developers simplify bitwise manipulation techniques in inline assembly and delete complex ones to reduce the potential for logic vulnerabilities arising from misunderstandings in the assembly code.
 
\textbf{S2: Adding Checks for Low-Level Calls in Inline Assembly.}
Developers utilize additional checks and assertions when making low-level calls in inline assembly, as these calls can bypass the logic in Solidity and introduce vulnerabilities via unauthorized or unintended behavior.

\textbf{S3:Removing Unnecessary Inline Assembly.}
 When unnecessary inline assembly is used in the contract, we have noticed that developers either delete it or minimize the logic in the inline assembly.

\textbf{S4: Invariant Validations in Upgradable Contracts.}
When performing contract upgrades, essential contract properties (i.e., invariants) must be preserved. These invariants include properties related to the contract such as interactions with external contracts, the contract state, and so on. Any inconsistency in these properties may affect the dependent logic and significantly impact the whole contract, resulting in vulnerabilities and unwanted behavior. This strategy includes using assert statements to ensure that these invariants are the same across all versions and that the functionality is consistent and unchanged.

\textbf{S5: Adding Validation to Ensure Withdrawals.}
This strategy involves developers adding validation checks to withdrawal functions within the contract logic. These checks verify if the withdrawal amount exceeds the balance of the recipient. If the withdrawal amount exceeds the balance, the transaction is reverted, and an appropriate error message is provided. Moreover, these checks are added to ensure that only authorized users can withdraw funds and that the withdrawal process is executed securely.

\textbf{S6: Adding Checks for Resetting Token Allowances.}
This strategy is employed by developers to validate token allowance management functions to prevent unintended changes to token allowances, reducing the risk of potential logic vulnerabilities such as unauthorized token transfers. By adding specific assertions, developers ensure that resetting token allowances aligns with the contract's intended functionality.

\textbf{S7: Adding Checks for Incorrect Fee Calculations.}
Developers modify the code to ensure accurate and reliable fee calculations by adding validation checks to fee calculation functions. This also includes adjusting the length and prefix calculations to ensure they are based on correct input values.

\textbf{S8: Adding Checks for Handling Zero Balances.}

Developers add checks to functions that handle token balances to ensure that zero balances are handled appropriately. These checks are added to manage cases where the balance of the recipient or the sender (msg.sender) is zero. Depending on the business logic of the contract, developers handle these cases differently, reducing the risk of potential vulnerabilities such as balance manipulation.

\textbf{S9: Optimizing and Refactoring Interest Rate Calculation.}
Refactoring the interest rate calculation involves avoiding unnecessary intermediate steps and redundant operations. For instance, combining the calculation of interest rates for borrowing and lending into a single calculation, while also adding proper checks for current interest rates and calculations.

\textbf{S10: Ensuring Consistency Between Function Modifiers and State Mutability Specifiers.}
 Addressing these types of logic vulnerabilities involves ensuring consistency between function modifiers and state mutability specifiers. This includes selecting the appropriate state mutability specifier (pure, view, payable, or nonpayable) based on the function's behavior and ensuring that the function modifier accurately reflects the function's interaction with external contracts or accounts.

\textbf{S11: Validating Selection of Operations Based on Opcode Values.}
When a developer decides to use operations in their logic using opcodes, they validate the selection of operations to ensure that opcode values fall within the expected range. This ensures that only authorized operations are executed, reducing the risk of potential vulnerabilities.

\textbf{S12: Accurate Parameter Handling in Function Invocations.}
We have noticed that some functionalities within the logic of the contract require certain inputs, either from inside or externally. Developers add validations for the parameters of such functions to ensure they adhere to expected formats, ranges, and constraints. Invalid or malicious inputs can lead to unexpected logic or vulnerabilities.

\textbf{S14: Minimizing and Validation of External Function Calls.}
External calls are checked by developers in two steps: first, by checking the return value to verify whether the operation succeeded or failed, and second, by using require statements or conditional checks to validate return values. For example, if calling an external contract to transfer tokens, a developer checks if the transfer was successful before proceeding with further operations.

\textbf{S15: Removing or Limiting Complex Data Structures.}
Complex data structures, such as nested arrays. They increase the likelihood of unintended behaviors in the contract logic. Developers remove or limit the complexity of these data structures. This mitigation strategy often entails refactoring the code to replace complex data structures with simpler alternatives, such as using flat arrays instead of nested structures. Developers also impose constraints on the depth or complexity of data structures to prevent them from becoming overly intricate.

 \section{Discussions and Implications}
\label{sec:empirical}
This section discusses the limitations and implications of our findings.

Our method has several limitations, primarily centered around the reliance on labeling through regular expressions, state-of-the-art tools, and manual labeling. Our approach includes multiple steps to identify vulnerabilities in both smart contracts and code changes. Initially, we utilize regular expressions, but this method often results in high false positives. To mitigate this, we focus on vulnerabilities that are also identified by state-of-the-art tools. However, not all vulnerabilities, especially those related to logic, can be effectively identified using regular expressions alone. Variants of vulnerability types that involve complex logic may remain undetected by this method. Furthermore, while state-of-the-art tools are robust, they may not encompass every type of vulnerability, particularly those originating from logical errors. Hence, our method includes labeling vulnerabilities based on the intersection of findings from both regular expressions and state-of-the-art tools. 

In essence, there may exist vulnerabilities that neither method detects, or new types of vulnerabilities that were previously unknown. Our manual checks have identified and defined new vulnerabilities. However, manually labeling a dataset as extensive as ours poses significant challenges, compounded by the absence of an automated method for double-checking identified vulnerabilities. Therefore, adopting a combined approach that integrates tools and regular expressions is necessary to effectively identify known vulnerabilities. To minimize this problem during testing and training the model, we have focused primarily on well-known logic vulnerabilities such as locked ether and incorrect equality sanity checks. 

The second limitation is capturing the context of vulnerabilities. The diff and selected lines describing vulnerabilities and their context in our work may have some limitations. There are instances where vulnerabilities depend on interactions across extensive portions of code, spanning large contracts or multiple libraries included in those contracts. Our model may not fully understand the implications of such widespread dependencies, as it was not trained on these specific scenarios. 

Furthermore, while we considered various types of vulnerabilities, the context in which they occur may not always be fully represented. Therefore, we focus on typical and crucial logic-related vulnerabilities defined in the literature for which we have confidence in our model's training on a substantial dataset. 
Moreover, our dataset is restricted to mitigations that developers have implemented in their code changes (diffs). This means that any unrecognized mitigations or ignored vulnerabilities remain unseen, creating gaps in our understanding of new vulnerabilities or mitigations. Therefore, we deliberately focused only on the mitigations and vulnerabilities that developers have fixed, similar to the approach used by Liu et al.~\cite{liu2018mining}. This method helps us avoid false positives by considering only the issues that developers have addressed, but it also means we might miss vulnerabilities that developers did not recognize, fix, or mitigate.

Finally, our model exhibits faster training times due to its architecture being a simplified version of the BERT architecture. However, it is important to note that its smaller size could make it more susceptible to overfitting compared to larger models. We anticipate that a larger dataset and more extensive training could enable larger models to achieve similar levels of accuracy.

\subsection{Implications}
\textbf{For Researchers. 
}Researchers can utilize Sóley to expand the scope of vulnerability detection in smart contracts. Utilizing minimal feature engineering techniques such as tokenization and slicing, our results show that Sóley and LLMs offer promising avenues for accuratly detecting diverse vulnerabilities. We believe additional exploring and the integration of control flows or data flows with LLMs could enhance vulnerability detection by integrating more advanced feature engineering. Adding more context, such as comments alongside code snippets or additional lines, could improve the semantic understanding of LLMs and enhance their detection capabilities. Moreover, Sóley can be used to curate datasets and generate labeled datasets with more accurate and semantically rich vulnerabilities. Future research could focus on developing automated approaches that utilize code changes to identify vulnerabilities in smart contracts at an early stage.  Lastly, there is a need for further exploration into automating the patching of smart contract vulnerabilities using code changes, as well as investigating the efficacy of different fix strategies commonly employed by developers.

\textbf{For Practitioners and Tool Builders.} Developing automated analysis tools that integrate Sóley with version control systems such as Git can improve detecting logic vulnerabilities in smart contracts.  Additionally, tool builders can extend existing analysis tools by integrating Sóley to detect logic-related vulnerabilities through plugins or extensions. Furthermore, tool builders can explore ways to augment their tools' capabilities to automate the patching of identified vulnerabilities. By providing developers with code suggestions for fixing their smart contracts, which may contribute to a more robust and accurate smart contracts. Leveraging LLMs and Sóley, practitioners can effectively assess, analyze, and secure smart contracts against potential vulnerabilities.

 \section{Related Work}
\label{sec:related_work}

This section describes related studies to our work, including empirical studies on logic code vulnerabilities in smart contracts, LLMs in smart contract vulnerability detection, and related datasets to our research.

\subsection{Empirical studies on Logic vulnerabilities in smart contracts}

In a study by Chaliasos et al.~\cite{chaliasos2024smart}, an empirical evaluation of existing tools for smart contract security was conducted. Their evaluation indicates that these tools often generate numerous insignificant reports, leading to an overwhelming number of false positives. Moreover, Chaliasos et al. highlight the inefficiency of these tools in detecting logic-related vulnerabilities, sanity checks, and logic errors. The study also shows that practitioners identify logic-related vulnerabilities and protocol layer vulnerabilities as significant threats that are not adequately addressed by existing security tools. 
Soud et al.~\cite{soud2022fly} defined logic vulnerabilities as inconsistencies with the contract and the programmer's intention. Among the identified logic-related vulnerabilities are Greedy Contract, which arises when the contract logic only locks Ether; Transaction Order Dependency, occurring when a contract's logic depends on the order of transaction execution within a block; Call to the Unknown vulnerability, where a function unexpectedly invokes the recipient's fallback function, potentially introducing malicious code; and the DoS by External Contract vulnerability. Zhang et al.~\cite{zhang2023demystifying} categorize 26 smart contract vulnerabilities into three groups. The first group includes hard-to-exploit or doubtful vulnerabilities. The second group is detectable by static analysis tools. The third group involves business logic vulnerabilities, such as price manipulation, which require high-level semantical oracles and are generally undetectable by current static analysis tools. Finally, Zhang et al~\cite{zhang2020framework} defined logic-related vulnerabilities as flaws in the decision logic, branching, sequencing, or
a computational algorithm, as found in natural language
specifications or implementation language. he study identified four primary logic-related categories in smart contracts: assembly code, DoS, fairness, and storage vulnerabilities.

\subsection{Detecting smart contract code vulnerabilities}

Detecting smart contract vulnerabilities has attracted significant attention recently. Several tools have emerged to analyze smart contract code. These tools can generally be categorized into two main groups: traditional static and dynamic analysis tools and tools based on machine learning and large language models (LLMs).

\textbf{Solidity vulnerability detection via traditional code analysis tools.
} Static analysis tools such as Slither~\cite{Slither} and smartcheck~\cite{Smartcheck} are designed to statically analyze smart contracts, detecting security vulnerabilities and bad coding patterns. Solhint~\cite{solhint} detects syntax-related vulnerabilities through static checks using a wide range of rules. Symbolic execution tools such as Mythril~\cite{Mythril} identify various vulnerabilities, including overflow/underflow and tx.origin issues. Additionally, Osiris~\cite{Osiris} facilitates integer vulnerability detection in Solidity by employing symbolic execution combined with taint analysis. Dynamic analysis tools such as Maian~\cite{nikolic2018finding} categorize vulnerable smart contracts into three main types: suicidal contracts, prodigal contracts, and greedy contracts. Maian analyzes Solidity contracts through dynamic analysis on a private blockchain to reduce the number of false positives
However, despite the effectiveness of these tools, Chaliasos et al.~\cite{chaliasos2024smart}'s empirical evaluation of these tools highlights that they do not adequately address logic-related vulnerabilities, which are often the root cause of high-impact attacks. Moreover, Zhang et al.~\cite{zhang2023demystifying} noted that more than 80\% of exploitable bugs remain undetectable by automated means.

\textbf{Solidity Vulnerability Detection using LLMs.}  

Sun et al.~\cite{sun2024gptscan} introduced GPTScan, a tool that combines GPT with static analysis. GPTScan prompts GPT to automatically recognize scenarios related to logic vulnerabilities in smart contracts. Additionally, GPT is trained to identify key variables and statements, which are then verified through static confirmation. 
Hu et al.~\cite{hu2023large} proposed GPTLENS based on GPT-4 as an auditor, analyzing smart contracts, and a critic, reviewing the audits. Their empirical findings indicate that GPTLENS brings significant enhancements compared to the traditional detection approaches. David et al.~\cite{david2023you} examined the feasibility of utilizing LLMs for smart contract security audits, evaluating them on 52 compromised Decentralized Finance (DeFi) smart contracts. Their findings indicate that GPT-4 and Claude models correctly identify vulnerabilities in 40\% of cases but exhibit a notable false positive rate, necessitating manual auditor involvement. Sun et al.~\cite{sun2023assbert} introduced ASSBert framework, designed for smart contract vulnerability classification based on active and semi-supervised bidirectional encoder representations from transformers (BERT) network. Finally, Jeon~\cite{jeon2021smartcondetect} proposed SmartConDetect, a static analysis tool designed to detect security vulnerabilities in Solidity-based smart contracts. Utilizing a pre-trained BERT model, SmartConDetect extracts code fragments from smart contracts and identifies vulnerable code patterns. 

Existing approaches mainly detect Solidity vulnerabilities, yet few are designed specifically to detect vulnerabilities within the logic of the contract. Although the nature of logic-related vulnerabilities, sanity checks, and inline assembly vulnerabilities, often making it more challenging to detect~\cite{chaliasos2024smart, chaliasos2022study}.

\subsection{Solidity Vulnerability Datasets}
A few research in the literature focused on datasets for Solidity vulnerabilities.
Durieux et al.~\cite{durieux2020empirical} curated a dataset with 69 annotated vulnerable Solidity contracts obtained from Etherscan. Their dataset includes labels specifying the location and category of the vulnerabilities. Moreover, The authors provided an unlabeled dataset containing 47,518 contracts. Soud et al.~\cite{soud2022automesc} constructed a dataset comprising approximately 6500 vulnerabilities and corresponding fix pairs. Chalias et al.~\cite{chaliasos2022study}  constructed a dataset of Solidity smart contracts containing 12.4M contracts. However, there are no available datasets specifically about logic vulnerabilities or logic-related inline assembly, nor a dataset with labeled vulnerabilities with code snippets of the vulnerable section.

 \section{Threats to Validity}
\label{sec:threats} 
This section addresses challenges to the internal, construct, and external validity. To identify validity threats in our study, we utilized the standard methodology proposed by Feldt et al.~\cite{feldt2010validity}.

\textbf{Internal Validity:} The labeling of vulnerabilities in large quantities poses a threat to internal validity due to the time and resource-intensive nature of the task. To mitigate the risk of falsely labeling contracts, we employed regular expressions and validated the results using two established tools, SmartCheck and Slither. Another internal validity concern relates to the availability of training data. Given the specificity of our task, obtaining a large amount of suitable data related to logic vulnerabilities was challenging. To address this, we selected vulnerabilities classified in the literature as logic vulnerabilities and compared them with other types of vulnerabilities.
The selection of models also presents a potential threat to internal validity. To mitigate this, we chose a range of well-reviewed LLMs. Additionally, there is a concern about the reliability of the quantitative analysis, considering the potential presence of unidentified vulnerabilities within the contracts. To address this, two experts reviewed the identified vulnerabilities from the manual labeling process. Furthermore, we have made the raw data openly accessible, allowing other researchers and users to validate the results.\\
\textbf{Construct Validity:} This could be the choice of evaluation metrics in our study. Nevertheless, we minimized this threat by selecting widely recognized metrics such as precision, recall, and F-measure. These metrics have been extensively employed in previous research, including the baseline. Another consideration regarding the construct validity of our study is the distribution of selected vulnerabilities, which may vary in quantity within the dataset. To ensure fair training and balanced data representation, we standardized the number of instances across all vulnerabilities, despite their varying occurrences in the dataset. The consistent high performance across all vulnerabilities suggests the efficacy of our approach. Additionally, to address concerns about training fairness across models, we maintained consistency by using identical hyperparameters, training datasets, optimizers, and the same machine for all models.\\
\textbf{External Validity:} A potential challenge to external validity is associated with the concern that the Ethereum smart contracts included in this study may not be a precise representation of all smart contracts across various blockchain platforms, such as Hyperledger Fabric or other contracts on different blockchains. Consequently, our quantitative analysis and approach may not accurately reflect the code changes or fixes in contracts beyond Ethereum smart contracts.  \section{Conclusion}
\label{sec:conclusion}
 Securing smart contracts has attracted many researchers and practitioners. One important aspect of improving smart contract security is addressing logic-related code vulnerabilities, which are the root cause of high-impact cyberattacks. In this work, we investigate logic-related vulnerabilities from code changes in smart contracts collected from GitHub. We introduce Sóley to detect these code vulnerabilities and compare its performance with other types of code vulnerabilities. Additionally, we explore and identify mitigation strategies for these code vulnerabilities based on developers' code changes. Sóley outperforms baselines and several LLMs in automatically identifying logic vulnerabilities. Interestingly, without requiring extensive feature engineering, Large Language Models (LLMs) performed effectively in this task, as evident from the results. In the future, we plan to explore the detection and fixing of other types of logic-related vulnerabilities. \section*{Data Availability}
Data is available upon request. 

\bibliographystyle{model1-num-names}

\appendix
\section{Regular Expressions}
Here, we list a subset of the regular expressions we used to extract certain type of vulnerabilities in smart contracts. 
\label{regex_list}
\small
\begin{description}
    \item[Vulnerability Name] Reentrancy (RE) send and transfer
    \item[Regex] \begin{verbatim}r'\b((?i)send|transfer)\s*\('
    \end{verbatim}

    \item[Vulnerability Name] Calls Loop (C-L)
    \item[Regex] \begin{verbatim}r'\b(?:uint|int|bool|address|bytes|mapping\s*\(|struct
    \s*\()(\s*(?!(?:memory|storage|calldata))\w+\s*)\s*;'
    \end{verbatim}

    \item[Vulnerability Name] Low-level Calls (L-L-C)
    \item[Regex] \begin{verbatim}r'\.call\s*\(.*\)'
    \end{verbatim}

    \item[Vulnerability Name] Locked Ether (L-E)
    \item[Regex] \begin{verbatim}r'\bfunction\s*\([\s\S]*payable\s{[\s\S]*?}(\v)[\s\S]?}'
    \end{verbatim}

    \item[Vulnerability Name] Incorrect Equality (I-E)
    \item[Regex] \begin{verbatim}r'\s*==\s*'
    \end{verbatim}

    \item[Vulnerability Name] Controlled Delegatecall
    \item[Regex] \begin{verbatim}r'\bdelegatecall\b'
    \end{verbatim}

    \item[Vulnerability Name] Timestamp Dependence
    \item[Regex] \begin{verbatim}r'\bnow\b|\b(block\.(?:timestamp|number|hash)\b)'
    \end{verbatim}

    \item[Vulnerability Name] Use of tx.origin
    \item[Regex] \begin{verbatim}r'\btx\.origin\b'
    \end{verbatim}
    
\end{description}
\appendix
\section{Vulnerability Distribution in the Dataset}
\begin{figure*}[ht] 
\centering
\includegraphics[width=0.40\textwidth]{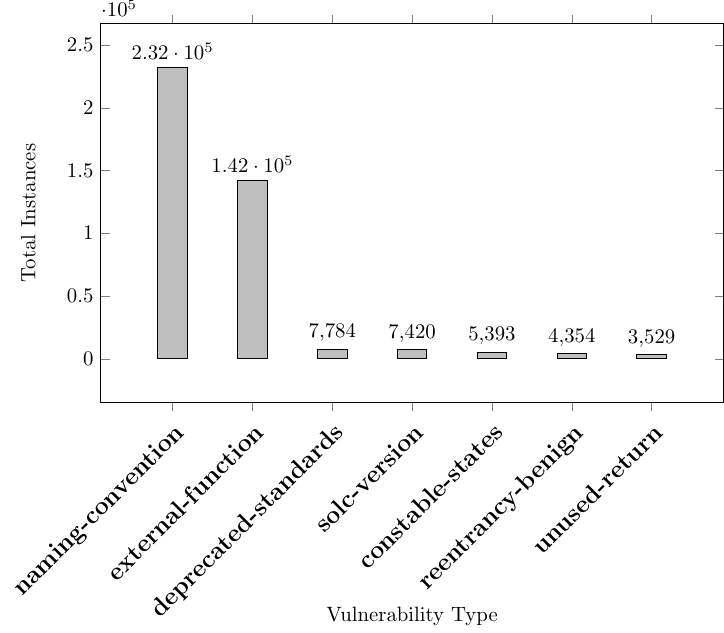}
       \label{fig:VT1111}
       \includegraphics[width=0.40\textwidth]{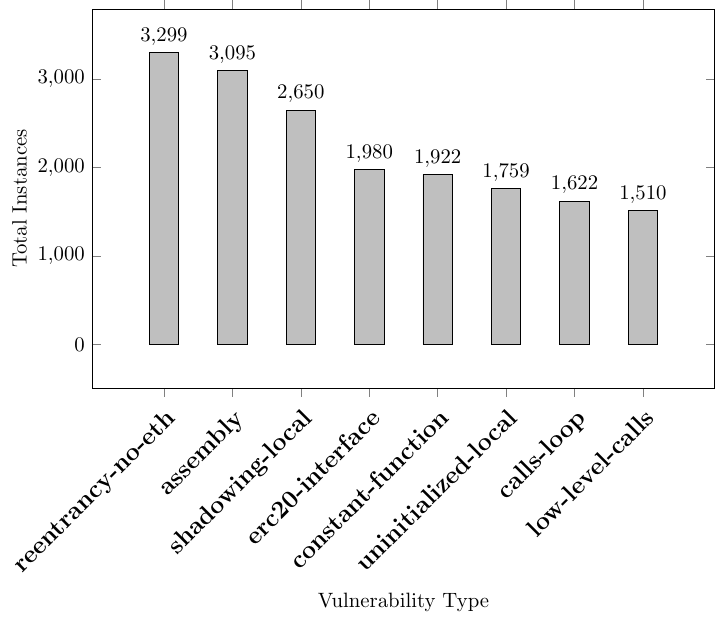}
       \label{fig:VT1}

\includegraphics[width=0.40\textwidth]{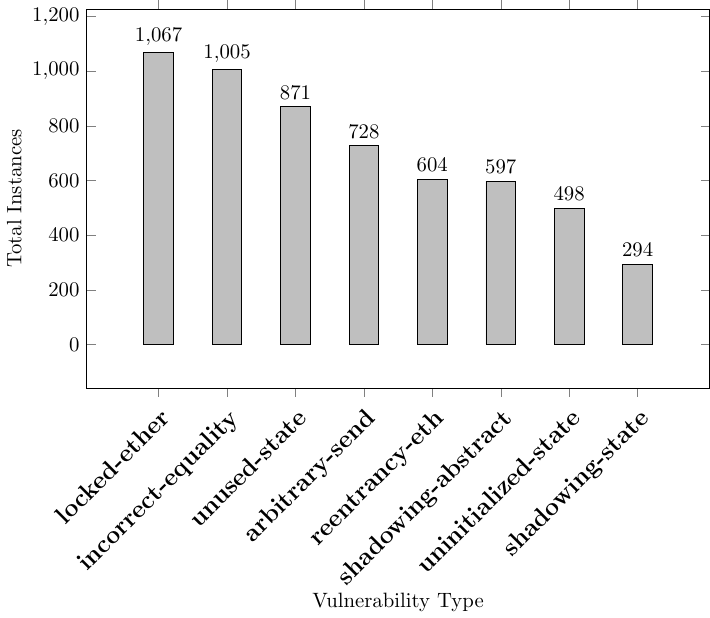}
           \label{fig:VT111}
       \includegraphics[width=0.40\textwidth]{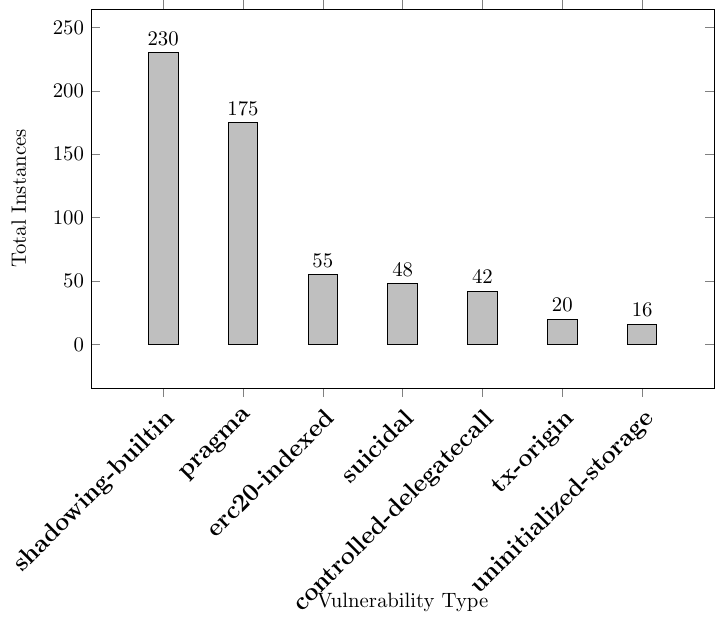}
       \label{fig:VT11}
       \caption{Vulnerability Distribution in Sóley Dataset}
        \label{fig:VT2}
\end{figure*}

In this section, we present the analysis of the vulnerability distribution within our dataset in Figure~\ref{fig:VT2}. 

\end{document}